\documentclass[11pt,a4paper]{article}
\usepackage{jcappub}
\usepackage{float}
\usepackage{rotating}
\usepackage{graphicx}
\usepackage{epsfig}
\usepackage{amsmath}
\usepackage{amssymb}
\usepackage{mathrsfs}  
\usepackage{subfigure}
\usepackage{time}
\usepackage{orcidlink}
\usepackage{relsize, comment}
\usepackage{xcolor}
\usepackage[utf8]{inputenc}

\usepackage{etoolbox}

\hypersetup{ linktoc=all,
    colorlinks=true, linkcolor={blue},  
       citecolor={red}, urlcolor={vividviolet}
}
\definecolor{rosy}{RGB}{230,235,252}
\definecolor{myframetitle}{RGB}{90,89,170}
\definecolor{myblocktitle}{RGB}{140,185,249}
\definecolor{mytitle}{RGB}{10,80,26}

\definecolor{darkgreen}{RGB}{27,130,45}
\definecolor{darkblue}{rgb}{0,0,0.3}
\definecolor{darkred}{rgb}{0.7,0,0}

\definecolor{light gray}{RGB}{220,220,220}
\definecolor{dark purple}{RGB}{108,0,217}
\definecolor{pink}{RGB}{190,20,100}
\definecolor{orang}{RGB}{193,63,0}
\definecolor{green}{RGB}{11,98,17}
\definecolor{darkpink}{RGB}{153,0,76}
\definecolor{bluegreen}{RGB}{0,102,102}
\definecolor{greenlagan}{RGB}{0,102,0}
\definecolor{redgreen}{RGB}{102,102,0}
\definecolor{Redgreen}{RGB}{153,76,0}
\definecolor{vividviolet}{rgb}{0.62, 0.0, 1.0}
\definecolor{amaranth}{rgb}{0.9, 0.17, 0.31}
\definecolor{palatinateblue}{rgb}{0.15, 0.23, 0.89}
\definecolor{brightpink}{rgb}{1.0, 0.0, 0.5}
\definecolor{cornflowerblue}{rgb}{0.39, 0.58, 0.93}
\definecolor{deepcarminepink}{rgb}{0.94, 0.19, 0.22}
\definecolor{radicalred}{rgb}{1.0, 0.21, 0.37}

\title{Prospective bounds on f(Q) gravity with pulsar timing arrays}

\author[a]{Mohammadreza Davari \orcidlink{0009-0001-8380-9328}}
\emailAdd{m.davari@khu.ac.ir}
\author[a]{Alireza Allahyari \orcidlink{0000-0002-4553-2436}}
\emailAdd{alireza.al@khu.ac.ir}

\affiliation[a]{Department of Astronomy and High Energy Physics, Kharazmi University, 15719-14911, Tehran, Iran \looseness=-1}

\abstract{
Pulsar timing arrays (PTAs) have recently provided compelling evidence for a stochastic gravitational wave background (SGWB) in the nanohertz frequency band, offering a unique window into fundamental physics. Here, we explore implications for symmetric teleparallel $f(Q)$ gravity, a theory in which deviations from General Relativity (GR) arise through the non-metricity scalar $f(Q)$. Crucially, tensor modes propagate at the speed of light in this framework. However, their amplitude undergoes a modified damping during their evolution. We adopt a model-independent parameterization and derive an analytic approximation to the tensor mode transfer function to obtain the spectral energy density of primordial inflationary gravitational waves. Comparison with the NANOGrav 15-year and IPTA second data releases show that the inferred damping parameter $n$ remains consistent with GR, yet allows small deviations that could be observable. We then conduct a Fisher information matrix forecasts which demonstrate that the Square Kilometre Array (SKA) observatory will improve these constraints by several orders of magnitude, offering the potential to distinguish $f(Q)$ gravity from GR with high precision. These results highlight PTAs as powerful probes of non-metricity-based modifications to gravity.}

\begin{document}
\maketitle
\flushbottom
\section{Introduction}
Recently, strong indications of the possibility of stochastic gravitational waves background in nHz have been obtained by various pulsar timing arrays (PTAs) observations \cite{NANOGrav:2023gor,Reardon:2023gzh,EPTA:2023fyk,Xu:2023wog}. But there is still no precise description for the source of these waves \cite{NANOGrav:2023hvm}. Focusing for concreteness on the NANOGrav \cite{NANOGrav:2023gor} signal, this stochastic GW background has a blue-tilted tensor spectrum, with the spectral index $n_T = 1.8 \pm 0.3$ \cite{Vagnozzi:2023lwo}. Modified gravity theories predict  different generation and propagation mechanisms for gravitational waves (GWs) and could generate a blue-tilted spectrum. Additionally, we need to consider the fact that the recent measurement of propagation speed of GWs from GW170817 relative to its electromagnetic counterpart GRB170817A \cite{LIGOScientific:2017zic} severely limits the deviations from the speed of light.

The $f(Q)$ gravity originates from the symmetric teleparallel framework \cite{Nester:1998mp,BeltranJimenez:2019esp}, where gravitation is described not by curvature or torsion but by the non-metricity scalar $Q$. The $f(Q)$ gravity \cite{Heisenberg:2023lru} has gained considerable popularity in the past couple of years and the bulk of the research efforts have been concentrated on cosmological applications~\cite{Lymperis:2022oyo, Paul:2022jnq, Narawade:2023tnn, Narawade:2023rip, Dimakis:2023uib}. This model has also been applied to large structure formation~\cite{Sokoliuk:2023ccw}, the development of relativistic versions of Modified Newtonian Dynamics (MOND)~\cite{Milgrom:2019rtd, DAmbrosio:2020nev}, bouncing cosmologies~\cite{Bajardi:2020fxh, Agrawal:2021rur, Gadbail:2023loj}, and even quantum cosmology~\cite{Dimakis:2021gby, Bajardi:2023vcc}. A lot of effort has also gone into constraining or testing $f(Q)$ models~\cite{Dialektopoulos:2019mtr, Ayuso:2020dcu, Barros:2020bgg, Frusciante:2021sio, Aggarwal:2022eae, De:2022jvo, Albuquerque:2022eac, Ferreira:2022jcd, Koussour:2023rly, Najera:2023wcw, Bouali:2023uik, Ferreira:2023tat, Subramaniam:2023old}. Extensions that involve incorporating boundary terms \cite{Capozziello:2023vne, De:2023xua, Paliathanasis:2023pqp} or non-minimally coupled scalar field \cite{Jarv:2018bgs, Harko:2018gxr} have also been explored.
Other very active area of research are black holes within $f(Q)$ gravity~\cite{Banerjee:2021mqk, Mustafa:2021bfs, Parsaei:2022wnu, Hassan:2022jgn, Hassan:2022ibc, Hassan:2022hcb, Venkatesha:2023tay, Jan:2023djj, Godani:2023nep, Javed:2023vmb, Mishra:2023bfe, Chanda:2022cod}, modified stellar solutions~\cite{Wang:2021zaz, Maurya:2022vsn, Maurya:2022wwa, Errehymy:2022gws, Sokoliuk:2022bwi, Calza:2022mwt, Bhar:2023yrf, Ditta:2023xhx, Maurya:2023szc}, and wormholes~\cite{Banerjee:2021mqk, Mustafa:2021bfs, Parsaei:2022wnu, Hassan:2022jgn, Hassan:2022ibc, Hassan:2022hcb, Venkatesha:2023tay, Jan:2023djj, Godani:2023nep, Mustafa:2023kqt, Mishra:2023bfe}. Also in this regard, some thought has been given to how observational data could be used to constrain $f(Q)$ gravity~\cite{Karmakar:2025yng, Arora:2025jxq, Sultanaa:2025ooz, Aggarwal:2025sqz, Najera:2025htf, Dutta:2025fqw, Mazumdar:2025vdk, Mandal:2025uht, Roy:2025nde, Paliathanasis:2025hjw, Garg:2025vyo, Bhattacharjee:2024txt, Wang:2024dkn, Mhamdi:2024xqd, Enkhili:2024dil, Wang:2024eai, Gadbail:2024als, Oliveros:2023mwl, Koussour:2023ulc, Maurya:2023muz, Goswami:2023knh, Myrzakulov:2023sir, Mussatayeva:2023aoa, Shi:2023kvu, Atayde:2023aoj, Gadbail:2023fjh, Koussour:2023gip, ElBourakadi:2023sch, Gadbail:2022hwq, Narawade:2022cgb, DAgostino:2022tdk, Pradhan:2022dml, Maurya:2022wwa, Koussour:2022irr, Lymperis:2022oyo, Khyllep:2022spx, Koussour:2022jss, Koussour:2022wbi, Koussour:2022nsc, Mandal:2021bpd, Atayde:2021pgb, Solanki:2021qni, Anagnostopoulos:2021ydo, Mandal:2021bfu, Barros:2020bgg, Lazkoz:2019sjl}.

In this paper, we assume that GW background has a primordial origin from an inflationary era with a modified damping term. To constrain the parameters of $f(Q)$ gravity using pulsar timing arrays data we apply a model-independent parameterization implemented in \cite{Capozziello:2022wgl}. In section \ref{2}, we introduce primordial GWs and their relation to PTAs. In section \ref{3}, we derive the approximate transfer function for tensor perturbations in $f(Q)$ gravity and find the spectral energy density. Then, in section \ref{4} by using NANOGrav 15-year data set (NG15) \cite{NANOGrav:2023gor} and International PTA second data release (IPTA2) \cite{Perera:2019sca,Antoniadis:2022pcn}, we constrain the model. Finally using Fisher formalism, we find the bounds from future PTAs data by Square Kilometre Array (SKA) observatory \cite{Weltman:2018zrl}. Section \ref{5} is summary and discussion.

\section{Primordial GWs}\label{2}

The evolution of linear, transverse-traceless perturbations for the tensor modes in GR are described by the following equation
\begin{equation}
	\ddot{h}_{ij} + 3 H \dot{h}_{ij} + ( k^2/a^2) h_{ij} = 0\,,
\end{equation}
where $H$ is the Hubble parameter. The relevant quantity in PTA observations is the spectral  energy density.
The spectral energy density of GWs can be derived as  \cite{Caprini:2018mtu}
\begin{equation}
	\label{spec}
	\Omega_{\text{GW}}(k,\eta) = \frac{1}{12H^2a^2}[T'(k,\eta)]^2 P_t(k)\,,
\end{equation}
where $T(k,\eta)$ is the transfer function. Transfer function describes the  evolution of GW modes after the modes re-enter the horizon. The quantity $P_t(k)$ is the primordial power spectrum of GWs at the end of the inflationary period. We may write this in terms of a tilt $n_t$ and a tensor amplitude $A$ as
\begin{equation}
	P_t(k) = A(\frac{k}{k_{\star}})^{n_t}\,,
\end{equation}
where 	$k_{\star}$ is a pivot scale set as $k_{\star}=0.05\,{\rm Mpc}^{-1}$. 
The transfer function in context of GR, is numerically computed in \cite{Zhao:2006mm}
\begin{equation}
	T(k,\eta_0)_{\rm GR} = \frac{3j_1(k\eta_0)}{k\eta_0} \sqrt{1.0 + 1.36(\frac{k}{k_{\rm eq}}) +2.50(\frac{k}{k_{\rm eq}})^2}\,,
\end{equation}
where $j_1$ is Bessel function and $\eta_0$ is the present conformal time. For GWs in PTA scales, we can use the approximation $(k \gg k_{\rm eq})$ and $[T'(k,\eta)]^2 = k^2 [T(k,\eta)]^2$ \cite{Vagnozzi:2023lwo}.

In pulsar timing experiments, it is convenient to express wavenumbers $k$ in terms of frequencies $f$ as \cite{Vagnozzi:2023lwo}  
\begin{eqnarray}
	f \simeq 1.54 \times 10^{-15} \left ( \frac{k}{{\rm Mpc}^{-1}} \right ) \,{\rm Hz}\,.
	\label{ktof}
\end{eqnarray}
We find that
$f_{\star} \simeq 7.7 \times 10^{-17}\,{\rm Hz}$. Also in PTAs, the GW spectral energy density is rather written in terms of the power spectrum of the GW strain $h_c$ given by
\begin{eqnarray}
	\Omega^{\rm PTA}_{\rm GW}(f) = \frac{2\pi^2}{3H_0^2}f^2h_c^2(f)\,.
	\label{eq:omegagwhc}
\end{eqnarray}
$h_c(f)$ is supposed to take a powerlaw form with respect to a reference frequency $f_{\rm yr}$ and can be expressed as
\begin{eqnarray}
	h_c(f) = A \left ( \frac{f}{f_{\rm yr}} \right ) ^{\alpha}\,,
	\label{eq:hc}
\end{eqnarray}
where $f_{\rm yr}=1\,{\rm yr}^{-1} \approx 3.17 \times 10^{-8}\,{\rm Hz}$. 
Finally using $\alpha = \frac{3-\gamma}{2}$,
the present GWs spectral energy density is given by \cite{Vagnozzi:2023lwo}
\begin{eqnarray}
	\Omega^{\rm PTA}_{\rm GW}(f) = A^2\frac{2\pi^2}{3H_0^2}\frac{f^{5-\gamma}}{{\rm yr}^{\gamma-3}}\,.
	\label{eq:omegagwhcpta}
\end{eqnarray}

\section{GWs in $f(Q)$ gravity}\label{3}
The $f(Q)$ gravity originates from the symmetric teleparallel framework \cite{Nester:1998mp,BeltranJimenez:2019esp}, where gravitation is described not by curvature or torsion but by the non-metricity scalar $Q$, whose action is given by
\begin{equation}
	S=\int  d^4x\, \sqrt{-g}\left[\dfrac{1}{2}f(Q)+\mathcal{L}_m\right]\,,
\end{equation}
where $\mathcal{L}_m$ is the matter field Lagrangian, and $g$ is the determinant of metric. Notice that, up to a total derivative, the above action and the Einstein-Hilbert one are equivalent for $f(Q)=Q$. The non-metricity scalar is given by \cite{BeltranJimenez:2019tme}
\begin{equation}
	Q = -Q_{\alpha \mu \nu} P^{\alpha \mu \nu} \,,
\end{equation}
where the non-metricity tensor $Q_{\alpha \mu \nu}$, and the non-metricity conjugate $P^{\alpha}{}_{\mu \nu}$, are given by
\begin{equation}
	Q_{\alpha \mu \nu} \equiv \nabla_{\alpha} g_{\mu \nu} \,.
\end{equation}
\begin{equation}
	P^{\alpha}{}_{\mu \nu} = - \frac{1}{2} L^\alpha{}_{\mu \nu} + \frac{1}{4} (Q^\alpha - \tilde{Q}^\alpha) - \frac{1}{4} \delta^\alpha_{(\mu} Q_{\nu)} \,,
\end{equation}
where $L^\alpha{}_{\mu \nu}$ is the disformation tensor, which takes the form
\begin{equation}
	L^\alpha{}_{\mu \nu} = \frac{1}{2} Q^\alpha{}_{\mu \nu} - Q_{(\mu \nu)}{}^\alpha \,,
\end{equation}
where $Q_\alpha$ and $\tilde{Q}_\alpha$ are two independent traces of the non-metricity, defined as
\begin{equation}
	Q_\alpha \equiv g^{\mu \nu} Q_{\alpha \mu \nu}, \quad \tilde{Q}_\alpha \equiv g^{\mu \nu} Q_{\mu \alpha \nu} \,.
\end{equation}
In the homogeneous and isotropic universe described by the FLRW metric, the non-metricity scalar is given by \cite{Capozziello:2022wgl}
\begin{equation}
	Q =6 H^2 = 6 \frac{\mathcal{H}^2}{a^2} \,.
\end{equation}

The evolution of linear, transverse-traceless perturbations for the tensor modes due to modifications of the gravity theory are generally described by the following equation \cite{Nishizawa:2017nef}
\begin{equation}\label{eom:general:gw:conformal}
	h''_{ij}+\left(2+\nu\right) \mathcal{H}  h'_{ij}+(c_T^2k^2 + a^2\mu^2) h_{ij}= a^2 \Gamma\gamma_{ij}\,,
\end{equation}
where prime denotes differentiation with respect to conformal time $\eta$. $\nu$ measures deviations from the standard amplitude damping, $c_T$ is the GWs propagation speed, $\mu$ an effective mass term, and $\Gamma\gamma_{ij}$ possible source contributions.

The propagation equation in $f(Q)$ gravity for each polarization $\lambda$ reads as~\cite{Karmakar:2025yng}
\begin{equation}\label{eom:fq:gw:conformal}
	h''_{(\lambda)}+2\mathcal{H}\left(1+\frac{d \log f_Q}{2\mathcal{H}d \eta}\right) h'_{(\lambda)}+k^2h_{(\lambda)}=0 \,,
\end{equation}
with $f_Q \equiv \partial f/\partial Q$.

$f(Q)$ models predict $c_T=1$, i.e. gravitational waves propagate at the speed of light, consistent with current observational constraints \cite{LIGOScientific:2017zic}. The modification adds  the damping parameter as
\begin{eqnarray}
	\nu&=& \frac{1}{\mathcal{H}} \frac{\mathrm{d} \log f_Q}{d \eta} \equiv \frac{1}{H}\frac{d \log f_Q}{d t} \,. 
\end{eqnarray}

The transfer function in modified gravity  can be approximated as a product of a correction factor and a relativity part \cite{Nunes:2018zot}. For $f(Q)$ gravity we have
\begin{equation}
	\label{transferMG}
	T(k,\eta)_{\text{MG}} = \mathrm{exp}\left(-\frac{1}{2} \int_{}^{\eta} \mathcal{H} \nu \, \mathrm{d}\eta'\right) T(k,\eta)_{\text{GR}}\,.
\end{equation}
We use a model-independent parameterization considering gravity mediated by non-metricity, with vanishing curvature and torsion; for describing the accelerated expansion of the universe with the function \cite{Capozziello:2022wgl}
\begin{equation}
	f(Q)=\alpha+\beta Q^{n}\,.
\end{equation}
We can see that one recovers GR for $\alpha=0$ and $\beta=1=n$, but for the $\Lambda$CDM model, we need $\alpha >0$ and $\beta=1=n$. Finally we find the transfer function as
\begin{equation}
	T(k,\eta)_{\text{MG}} = \left(\frac{\mathcal{H}(a) a_e}{\mathcal{H}(a_e) a}\right)^{1-n} T(k,\eta)_{\text{GR}}\,,
\end{equation}
where $a_e$ refers to the scale factor at horizon entry.
We consider the evolution of background in $\Lambda$CDM model \cite{Khyllep:2021pcu,Ferreira:2022jcd}; So the Hubble parameter can be written as
\begin{equation}
	\mathcal{H}^2(a)=a^2 \mathcal{H}_{0}^{2}\Big[\Omega_{m,0}\,a^{-3}+\Omega_{r,0}\,a^{-4}+\Omega_{\Lambda,0}\,\Big] \,,
\end{equation}
with matter density parameter $\Omega_{m,0} = 0.315$, radiation density parameter $\Omega_{r,0} = 9 \times 10^{-5}$, dark energy density parameter $\Omega_{\Lambda,0} = 0.68491$ and $\mathcal{H}_{0} = 67.36$ km/s/Mpc from Planck 2018 \cite{Planck:2018vyg}. Also the scale factor at horizon entry is related to frequency as
\begin{equation}
	a_e \approx \frac{3.33870198 \times 10^{-21}}{f}
	\,,
\end{equation}
Using the above equations we obtain the spectral energy density of gravitational waves at present time as
\begin{equation}
	\begin{split}
		\Omega_{\text{GW}}(f) = A^2 \frac{2\pi^2}{3H_0^2} \left(\frac{f^{5-\gamma}}{\text{yr}^{\gamma-3}}\right)
		\left(\frac{\mathcal{H}_0 a_e}{\mathcal{H}_e}\right)^{2-2n} \\
		\times \left(1 + \frac{\mathcal{H}_0 (1-n)(\Omega_{m,0}+\Omega_{r,0})}{4\pi f} \right)^2
		\,.
	\end{split}
\end{equation}

\section{Bounds by PTAs}\label{4}
In this section, we use NANOGrav 15 year data (NG15) and IPTA second data release (IPTA2) to constrain the parameter $n$. The parameters $\alpha$ and $\beta$ do not enter the final expression for spectral energy density. In the following, we show that more future data with improved precision can  enhance the accuracy of the results.

\subsection{NANOGrav and IPTA}
\begin{figure}[htbp]
	\centerline{\includegraphics[scale=0.8]{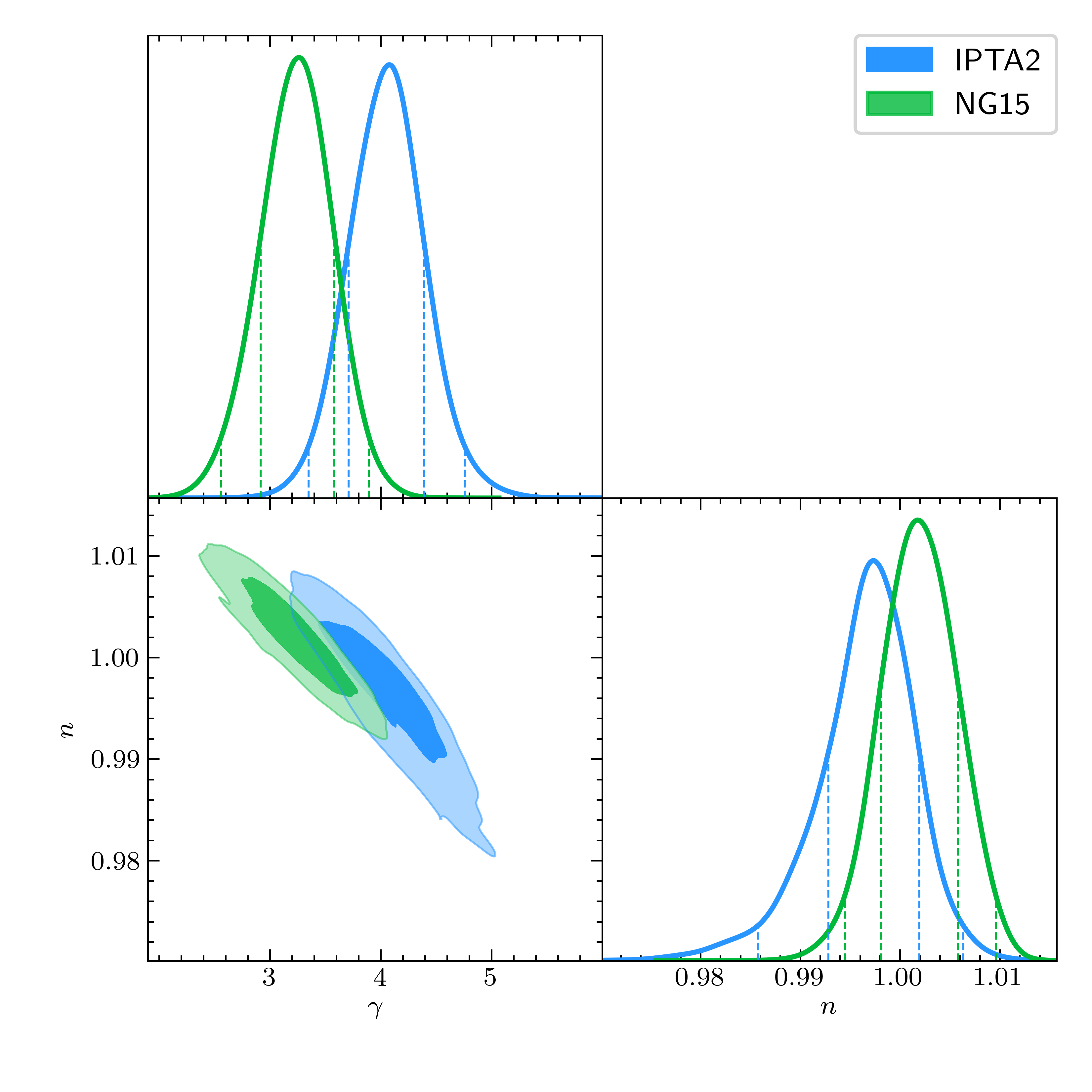}}
	\caption{The posterior plots and marginal posteriors of $f(Q)$ gravity parameters with $\text{log}_{10}A = -14.2$. The contours show at 68\% and 95\%  confidence levels for International PTA second data release (IPTA2) in blue and NANOGrav 15-year data set (NG15) in green.}
	\label{fig_fq}
\end{figure}

\begin{figure}[htbp]
	\centerline{\includegraphics[scale=1.1]{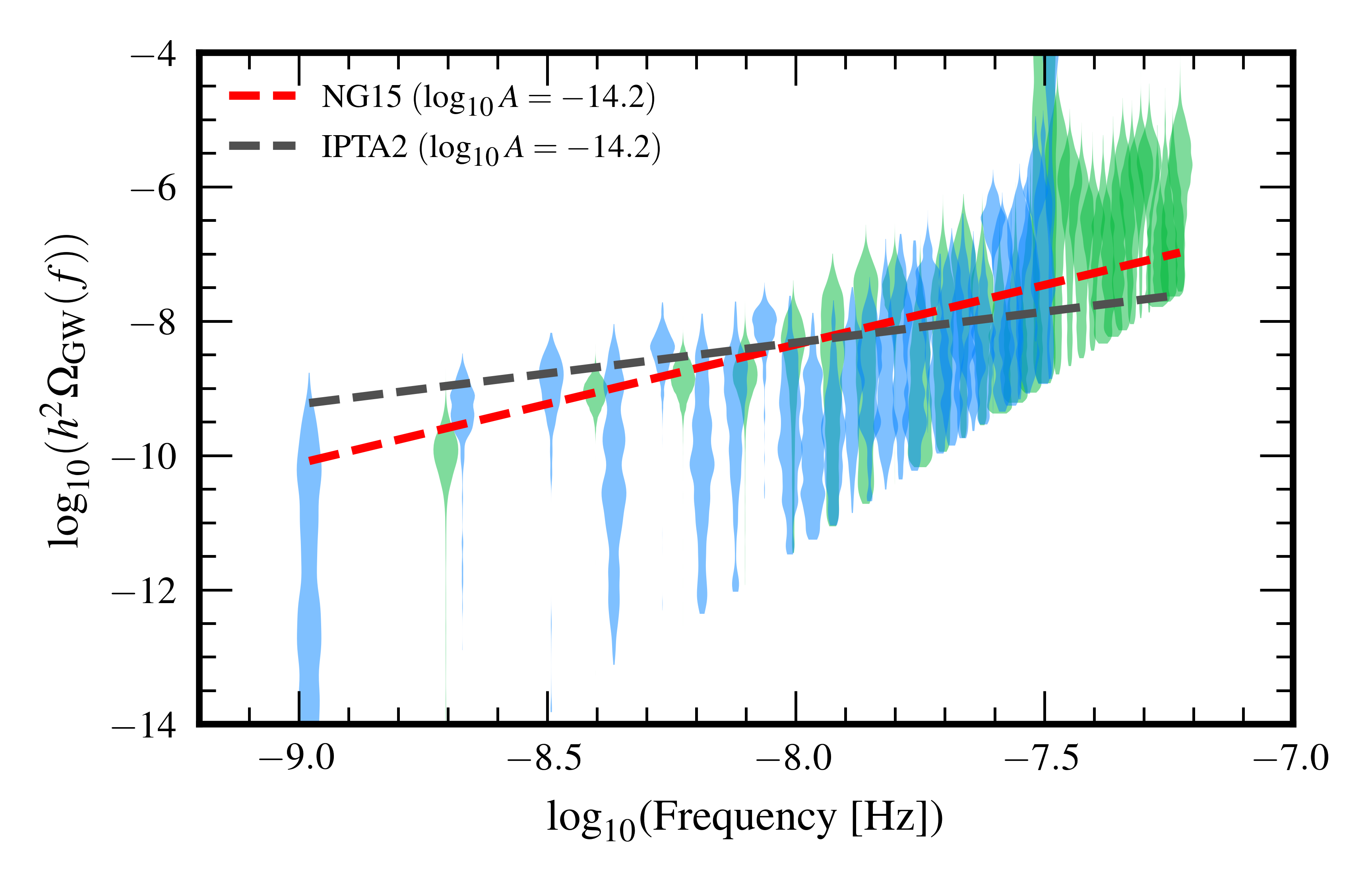}}
	\caption{The current spectral energy density of GWs as a function of frequency in logarithmic scales for $f(Q)$ model. The violin plots show NANOGrav 15-year data set (NG15) and International PTA second data release (IPTA2).}
	\label{omega}
\end{figure}

We use the python package  \texttt{PTArcade} \cite{Mitridate:2023oar}, that provides an accessible way to perform Bayesian analyses with PTAs data. We consider uniform priors on the parameters as $(0<\gamma<6)$ and $(0<n<2)$. We fixed the parameter $\text{log}_{10}A$ with three values $\text{log}_{10}A = -13.6, -14.2, -14.8$, because it is hardly constrained. The results are summarized in table \ref{table1}. Also the results in case of $\text{log}_{10}A = -14.2$ are illustrated in figure \ref{fig_fq}. The green lines show NG15 and blue lines show IPTA2. As we see for red-tilt spectrum we often find $n < 1$, and for blue-tilt spectrum we often find $n > 1$. Also upper amplitude gives $n < 1$ and lower amplitude gives $n > 1$. Predictions  are consistent with GR given uncertainties. This shows that pulsar timing arrays data is a very powerful tool for studying and constraining modified gravity theories.
In figure \ref{omega}, for concreteness, we show $h^2\Omega_{\text{GW}}$ as a function of frequency in logarithmic scales using the best fit of values of the parameters obtained from NG15 and IPTA2.

\begin{table}
	\centering
	\begin{tabular}{cccc}
		\hline
		Parameter &  NG15 & IPTA2 & $\text{log}_{10}A$ (fixed) \\
		\hline
		
		$n$ & $0.9857\pm 0.0039$ & $0.9802^{+0.0055}_{-0.0037}$ & $-13.6$ \\
		
		$\gamma$ & $3.14\pm 0.34$ & $3.96\pm 0.36$ & $-13.6$ \\
		\hline
		$n$ & $1.0019\pm 0.0039$ & $0.9965^{+0.0055}_{-0.0037}$ & $-14.2$ \\
		
		$\gamma$ & $3.24\pm 0.34$ & $4.06\pm 0.36$ & $-14.2$ \\
		\hline
		$n$ & $1.0181\pm 0.0039$ & $1.0127^{+0.0055}_{-0.0037}$ & $-14.8$ \\
		
		$\gamma$ & $3.34\pm 0.34$ & $4.15\pm 0.36$ & $-14.8$ \\
		\hline
	\end{tabular}
	\caption{Constraints at 68\% confidence level from NG15 and IPTA2.}
	\label{table1}
\end{table}

\subsection{SKA observatory}
\begin{figure}[htbp]
	\centerline{\includegraphics[scale=0.6]{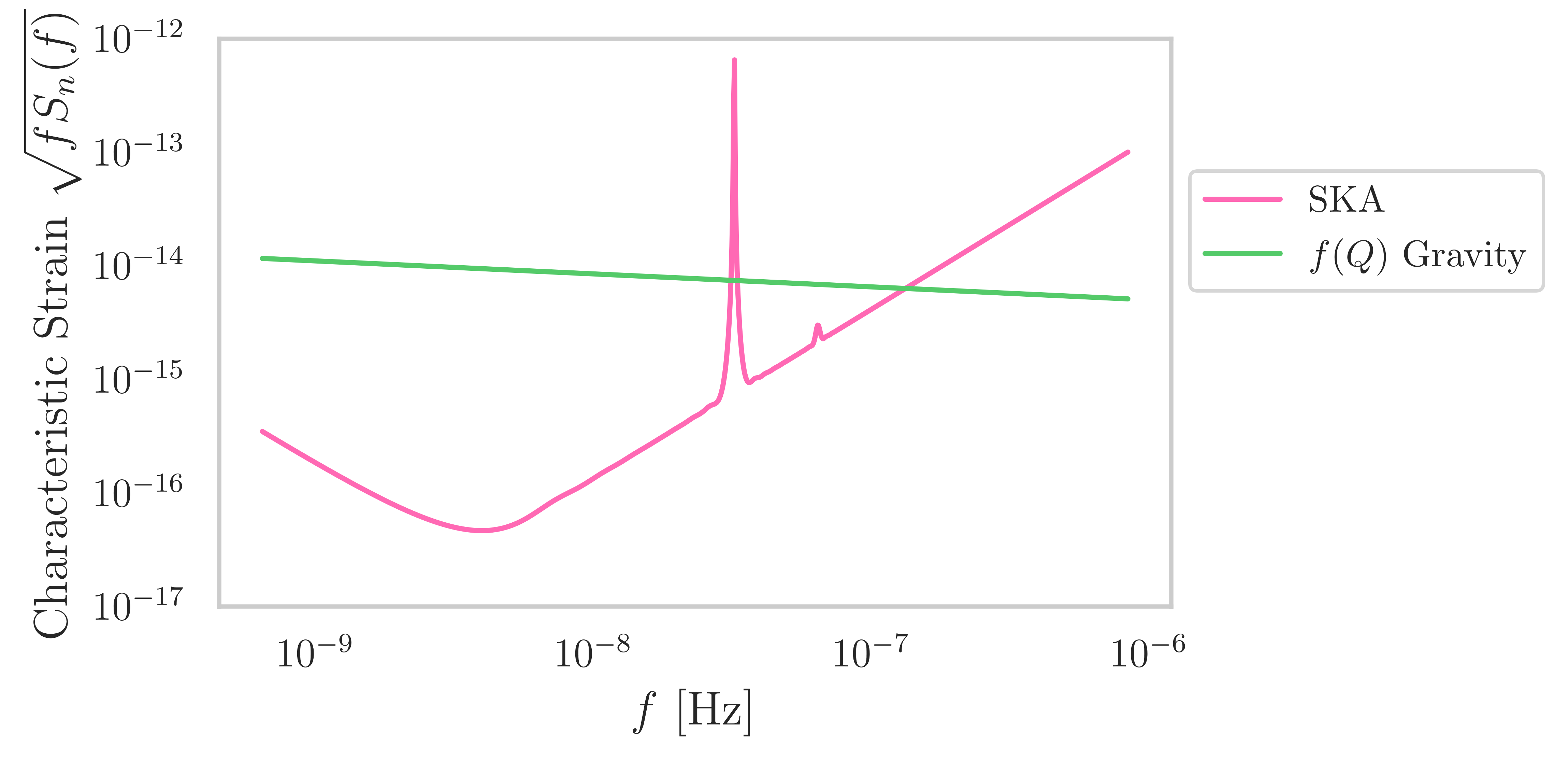}}
	\caption{The gravitational wave signal spectrum from $f(Q)$ gravity (green) compared to the SKA noise spectrum (pink).}
	\label{figsig}
\end{figure}

In this section we show that improving the quantity and precision of PTAs data results in stronger parameter constraints. The predictions are obtained by constructing the Fisher information matrix from estimated SKA observations. We first compare the model-derived gravitational wave signal spectrum with the SKA noise spectrum using the python package \texttt{SGWBProbe} \cite{Campeti:2020xwn}. As shown in Figure \ref{figsig}, the SKA is sensitive enough to detect the signal predicted by the model. The signal is written as \cite{Hazboun:2019vhv}
\begin{equation}
	S_h(f) = \Omega_{\text{GW}}(f) \frac{3 H_0^2}{2 \pi^2 f^3} \,.
\end{equation}

The Fisher Information Matrix (FIM) for two pairs of pulsars I,L and J,K is written as \cite{Babak:2024yhu}
\begin{equation}
	F_{\alpha \beta} 
	\equiv 
	\sum_{f_k}
	C^{-1}_{IJ} C^{-1}_{KL} 
	\frac{\partial (R_{JK} S_h)}{\partial \theta^\alpha}
	\frac{\partial (R_{LI} S_h)}{\partial \theta^\beta} \,,
\end{equation}
where $\theta^{\alpha}$ are model parameters, $f_k = \frac{k}{T_{\text{obs}}}$, $C$ is signal covariance matrix includes gravitational waves term $C_h$ and noise term $C_n$ as
\begin{equation}
	C_{IJ}=C_{n,IJ}+C_{h,IJ}\,,
\end{equation}
\begin{equation}    
	C_{n,IJ}
	= \delta_{IJ} P_{I}^{WN}\,,
\end{equation}
\begin{equation}
	C_{h,IJ}    	 
	= R_{IJ} S_h(f)\,,
\end{equation}
we only consider white noise $P_{I}^{WN}$ and ignore other noises
\begin{equation}
	P_I^{\rm WN} =  2 \sigma^2 \Delta t \,,
\end{equation}
where $\Delta t$ is the inverse of the observing cadence and $\sigma$ is the rms timing uncertainty. Also $R_{IJ}$ introduced as
\begin{equation}
	R_{IJ} = 
	\chi_{IJ} \cdot  {\cal R}(f)  
	\left [
	{\cal T}_I(f)
	{\cal T}_J(f)
	{T_{IJ} }/{T_\text{obs} }
	\right]^{1/2} ,
\end{equation}
\begin{equation}
	\chi_{IJ} 
	=
	\frac{1}{2} +
	\frac{3}{2} \xi_{IJ}
	\left[\ln\xi_{IJ} - \frac{1}{6}\right]
	+\frac{1}{2}\,\delta_{IJ}\,,
	\label{e:HD}
\end{equation}
\[\xi_{IJ} \equiv 
\left({1-\cos(\zeta_{IJ})}\right)/{2} \,, \]
where the frequency dependent, sky-averaged, quadratic response function defined as ${\cal R}(f) \equiv 1/12 \pi^2 f^2$ and the transfer function ${\cal T}(f) \simeq \left[  1+ 1/\left (f T_\text{\tiny obs} \right ) \right ]^{-6}$. The first geometrical factor $\chi_{IJ}$ is the well-known Hellings-Downs correlation pattern as a function of the angular separation $\zeta_{IJ}$ between pulsars as required by symmetries \cite{Kehagias:2024plp}. And effective overlapping time defined as $T_{IJ}={\rm min}[T_{I},T_{J}] $. For additional information see \cite{Babak:2024yhu}.

\begin{figure}[htbp]
	\centerline{\includegraphics[scale=0.4]{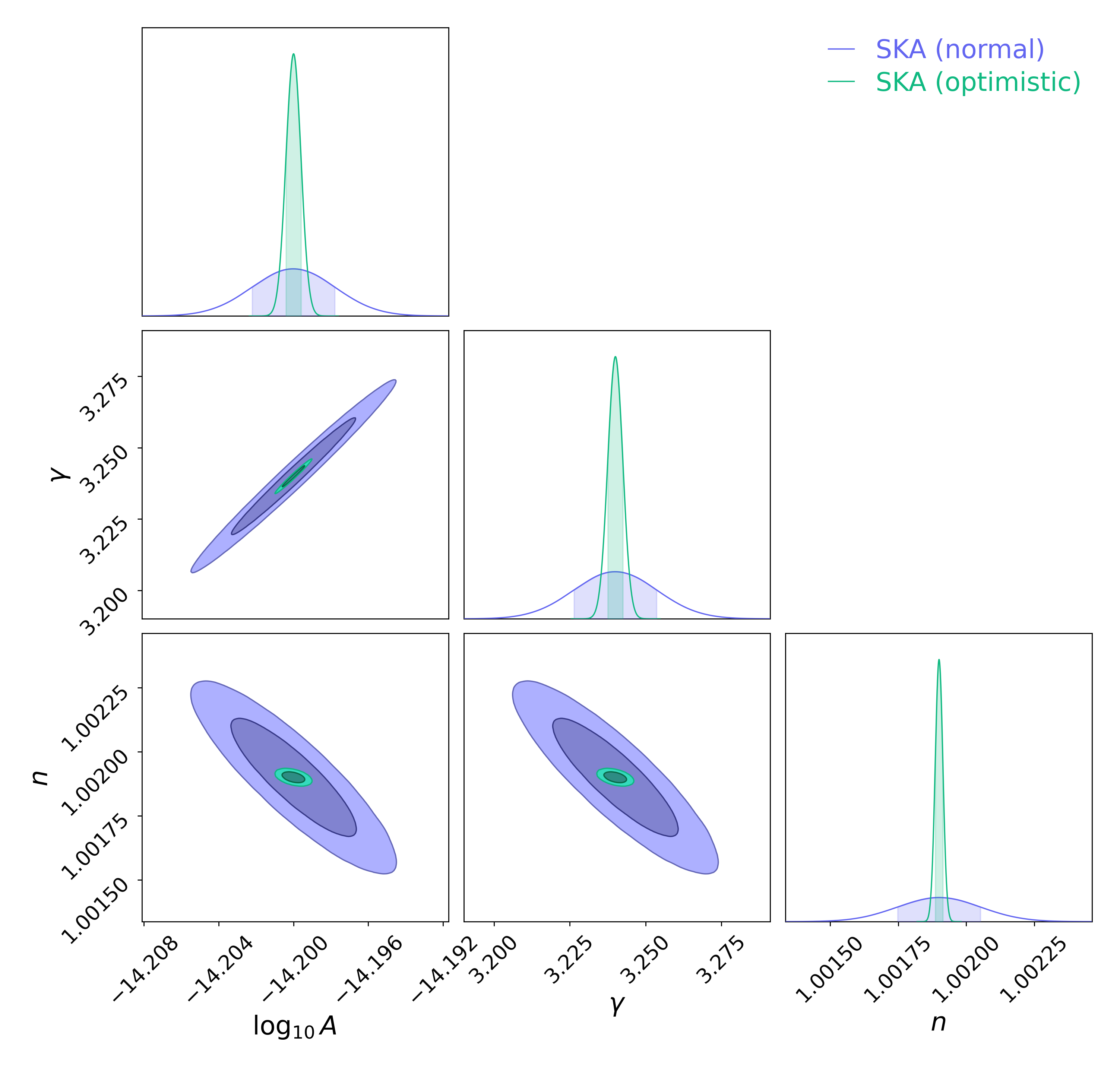}}
	\caption{The posterior plots and marginal posteriors of the parameters. The contours show at 68\% and 95\%  confidence levels for two scenarios SKA (normal) in indigo and SKA (optimistic) in emerald.}
	\label{figska}
\end{figure}

Finally, the covariance matrix $C_{\alpha \beta}$, is obtained by inverting the FIM, from which one can estimate uncertainties as $\sigma_\alpha \equiv \sqrt{F_{\alpha \alpha}^{-1}}$. In this calculation, we consider the two scenarios performed in \cite{Lee:2021zqw}, which are listed in table \ref{tableska}.
\begin{table}
	\centering
	\begin{tabular}{ccccc}
		\\
		\hline
		SKA &  $N_{\text{p}}$ & $T_{\text{obs}}$[yr] & $\Delta t$[week] & $\sigma$[ns] \\
		\hline
		{Normal} & 200 & 20 & 2 & 50 \\
		Optimistic & 1000 & 30 & 1 & 10 \\
	\end{tabular}
	\caption{PTA parameters considered for SKA. Here $N_{\text{p}}$ is the number of pulsars, $T_{\text{obs}}$ is the observation time, $\Delta t$ is the cadence and $\sigma$ is the root-mean-square timing residuals.}
	\label{tableska}
\end{table}
We set the mean values of the parameters from the best fit of NG15. For the convenience of calculating the inverse of the covariance matrix, we assume that there are $N$ groups of 100 pulsars ($N_{\text{p}} = 100 N$). So we have
\begin{equation}
	F_{\alpha \beta} 
	\equiv 
	N \sum_{f_k}
	C^{-1}_{IJ} C^{-1}_{KL} 
	\frac{\partial (R_{JK} S_h)}{\partial \theta^\alpha}
	\frac{\partial (R_{LI} S_h)}{\partial \theta^\beta} \,.
\end{equation}
It should be noted that we have simulated pulsars at uniform random angles in the sky. The results are listed in table \ref{resska} and illustrated in figure \ref{figska}. As we see, future PTAs data from SKA can determine the parameters with higher precision.

\begin{table}
	\centering
	\begin{tabular}{ccc}
		\hline
		Parameter &  SKA (normal) & SKA (optimistic)  \\
		\hline
		$n$ & $1.00191^{+0.00014}_{-0.00016}$ & $1.0019\pm 0.000014$  \\
		
		$\gamma$ & $3.24\pm 0.014$ & $3.24^{+0.0025}_{-0.0024}$  \\
		
		$\text{log}_{10}A$ & $-14.2\pm 0.0022$ & $-14.2^{+0.00041}_{-0.00040}$ \\
		\hline
		
	\end{tabular}
	\caption{Constraints at 68\% confidence level from SKA.}
	\label{resska}
\end{table}

The Fisher matrix forecasts presented here assume an idealized PTA noise model and are intended to illustrate the sensitivity of SKA-era observations to the modified gravity parameter $n$. In our analysis, $n$ affects the frequency dependence of the stochastic gravitational wave background through a modified damping of tensor modes, leading to constraints that are driven primarily by the spectral shape rather than
the overall amplitude.

We have only considered white timing noise. However, in realistic PTA analyses, additional noise contributions such as intrinsic red timing noise or common-spectrum processes may be present and could moderately broaden the inferred parameter uncertainties.
As usual, Fisher matrix estimates correspond to lower bounds on parameter uncertainties under the assumption of a Gaussian likelihood. A full Bayesian analysis of future PTA data is therefore expected to yield somewhat larger credible intervals, while preserving the qualitative sensitivity to deviations from GR.

\section{Summary and discussion} \label{5}
In this work, we have explored the constraints of recent pulsar timing array (PTA) observations for symmetric teleparallel $f(Q)$ gravity. Adopting a model-independent parameterization, we focused on the modified damping of tensor modes encoded in the parameter $n$ while ensuring that gravitational waves propagate at the speed of light, consistent with multi messenger constraints from GW170817. Using an analytic approximation for the tensor mode transfer function, we derived the present-day spectral energy density of primordial inflationary gravitational waves in the context of $f(Q)$ gravity. Our analysis with NANOGrav 15-year (NG15) and International PTA second data release (IPTA2) sets tight constraints on $n$, which remains consistent with the GR value of $n=1$. Nonetheless, the results allow small deviations, especially for blue-tilted tensor spectra or lower amplitudes, indicating that future, more sensitive observations may reveal subtle non-metricity effects. The posterior distributions demonstrate the potential of PTAs to probe modifications of gravity beyond the standard $\Lambda$CDM framework.

Our Fisher matrix forecasts for the Square Kilometre Array (SKA) show that future PTA data could improve constraints on $f(Q)$ parameters. In the optimistic scenario, uncertainties on $n$ shrink to $\mathcal{O}(10^{-5})$, making it feasible to distinguish $f(Q)$ gravity from GR with more precision. This highlights the transformative role of next-generation PTAs in testing fundamental physics and constraining non-metricity based deviations from Einstein’s theory. Overall, our results underline that current PTAs already provide meaningful bounds on $f(Q)$ gravity, and future observations will enable a high precision exploration of non-metricity effects in the propagation of gravitational waves.


\bibliographystyle{JHEP}
\bibliography{ref}

@article{Caprini:2018mtu,
	author = "Caprini, Chiara and Figueroa, Daniel G.",
	title = "{Cosmological Backgrounds of Gravitational Waves}",
	eprint = "1801.04268",
	archivePrefix = "arXiv",
	primaryClass = "astro-ph.CO",
	doi = "10.1088/1361-6382/aac608",
	journal = "Class. Quant. Grav.",
	volume = "35",
	number = "16",
	pages = "163001",
	year = "2018"
}

@article{Zhao:2006mm,
	author = "Zhao, Wen and Zhang, Yang",
	title = "{Relic gravitational waves and their detection}",
	eprint = "astro-ph/0604458",
	archivePrefix = "arXiv",
	doi = "10.1103/PhysRevD.74.043503",
	journal = "Phys. Rev. D",
	volume = "74",
	pages = "043503",
	year = "2006"
}

@article{Vagnozzi:2023lwo,
	author = "Vagnozzi, Sunny",
	title = "{Inflationary interpretation of the stochastic gravitational wave background signal detected by pulsar timing array experiments}",
	eprint = "2306.16912",
	archivePrefix = "arXiv",
	primaryClass = "astro-ph.CO",
	doi = "10.1016/j.jheap.2023.07.001",
	journal = "JHEAp",
	volume = "39",
	pages = "81--98",
	year = "2023"
}

@article{Nishizawa:2017nef,
	author = "Nishizawa, Atsushi",
	title = "{Generalized framework for testing gravity with gravitational-wave propagation. I. Formulation}",
	eprint = "1710.04825",
	archivePrefix = "arXiv",
	primaryClass = "gr-qc",
	doi = "10.1103/PhysRevD.97.104037",
	journal = "Phys. Rev. D",
	volume = "97",
	number = "10",
	pages = "104037",
	year = "2018"
}

@article{Karmakar:2025yng,
	author = "Karmakar, Purnendu and Haridasu, Sandeep",
	title = "{Dynamical Dark Energy or Modified Gravity? Signatures in Gravitational Wave Propagation}",
	eprint = "2509.07976",
	archivePrefix = "arXiv",
	primaryClass = "gr-qc",
	month = "9",
	year = "2025"
}

@article{Nunes:2018zot,
	author = "Nunes, Rafael C. and Alves, Marcio E. S. and de Araujo, Jose C. N.",
	title = "{Primordial gravitational waves in Horndeski gravity}",
	eprint = "1811.12760",
	archivePrefix = "arXiv",
	primaryClass = "gr-qc",
	doi = "10.1103/PhysRevD.99.084022",
	journal = "Phys. Rev. D",
	volume = "99",
	number = "8",
	pages = "084022",
	year = "2019"
}

@article{Capozziello:2022wgl,
	author = "Capozziello, Salvatore and D'Agostino, Rocco",
	title = "{Model-independent reconstruction of f(Q) non-metric gravity}",
	eprint = "2204.01015",
	archivePrefix = "arXiv",
	primaryClass = "gr-qc",
	doi = "10.1016/j.physletb.2022.137229",
	journal = "Phys. Lett. B",
	volume = "832",
	pages = "137229",
	year = "2022"
}

@article{Planck:2018vyg,
	author = "Aghanim, N. and others",
	collaboration = "Planck",
	title = "{Planck 2018 results. VI. Cosmological parameters}",
	eprint = "1807.06209",
	archivePrefix = "arXiv",
	primaryClass = "astro-ph.CO",
	doi = "10.1051/0004-6361/201833910",
	journal = "Astron. Astrophys.",
	volume = "641",
	pages = "A6",
	year = "2020",
	note = "[Erratum: Astron.Astrophys. 652, C4 (2021)]"
}

@article{Mitridate:2023oar,
	author = {Mitridate, Andrea and Wright, David and von Eckardstein, Richard and Schr{\"o}der, Tobias and Nay, Jonathan and Olum, Ken and Schmitz, Kai and Trickle, Tanner},
	title = "{PTArcade}",
	eprint = "2306.16377",
	archivePrefix = "arXiv",
	primaryClass = "hep-ph",
	reportNumber = "FERMILAB-PUB-23-588-T",
	month = "6",
	year = "2023"
}

@article{Babak:2024yhu,
	author = "Babak, Stanislav and Falxa, Mikel and Franciolini, Gabriele and Pieroni, Mauro",
	title = "{Forecasting the sensitivity of pulsar timing arrays to gravitational wave backgrounds}",
	eprint = "2404.02864",
	archivePrefix = "arXiv",
	primaryClass = "astro-ph.CO",
	reportNumber = "CERN-TH-2024-039",
	doi = "10.1103/PhysRevD.110.063022",
	journal = "Phys. Rev. D",
	volume = "110",
	number = "6",
	pages = "063022",
	year = "2024"
}

@article{Kehagias:2024plp,
	author = "Kehagias, Alex and Riotto, Antonio",
	title = "{The PTA Hellings and Downs correlation unmasked by symmetries}",
	eprint = "2401.10680",
	archivePrefix = "arXiv",
	primaryClass = "gr-qc",
	doi = "10.1088/1475-7516/2024/06/059",
	journal = "JCAP",
	volume = "06",
	pages = "059",
	year = "2024"
}

@article{Lee:2021zqw,
	author = "Lee, Vincent S. H. and Taylor, Stephen R. and Trickle, Tanner and Zurek, Kathryn M.",
	title = "{Bayesian Forecasts for Dark Matter Substructure Searches with Mock Pulsar Timing Data}",
	eprint = "2104.05717",
	archivePrefix = "arXiv",
	primaryClass = "astro-ph.CO",
	reportNumber = "CALT-TH-2021-016",
	doi = "10.1088/1475-7516/2021/08/025",
	journal = "JCAP",
	volume = "08",
	pages = "025",
	year = "2021"
}

@article{LIGOScientific:2017zic,
	author = "Abbott, B. P. and others",
	collaboration = "LIGO Scientific, Virgo, Fermi-GBM, INTEGRAL",
	title = "{Gravitational Waves and Gamma-rays from a Binary Neutron Star Merger: GW170817 and GRB 170817A}",
	eprint = "1710.05834",
	archivePrefix = "arXiv",
	primaryClass = "astro-ph.HE",
	reportNumber = "LIGO-P1700308",
	doi = "10.3847/2041-8213/aa920c",
	journal = "Astrophys. J. Lett.",
	volume = "848",
	number = "2",
	pages = "L13",
	year = "2017"
}

@article{Campeti:2020xwn,
	author = "Campeti, Paolo and Komatsu, Eiichiro and Poletti, Davide and Baccigalupi, Carlo",
	title = "{Measuring the spectrum of primordial gravitational waves with CMB, PTA and Laser Interferometers}",
	eprint = "2007.04241",
	archivePrefix = "arXiv",
	primaryClass = "astro-ph.CO",
	doi = "10.1088/1475-7516/2021/01/012",
	journal = "JCAP",
	volume = "01",
	pages = "012",
	year = "2021"
}

@article{NANOGrav:2023gor,
	author = "Agazie, Gabriella and others",
	collaboration = "NANOGrav",
	title = "{The NANOGrav 15 yr Data Set: Evidence for a Gravitational-wave Background}",
	eprint = "2306.16213",
	archivePrefix = "arXiv",
	primaryClass = "astro-ph.HE",
	doi = "10.3847/2041-8213/acdac6",
	journal = "Astrophys. J. Lett.",
	volume = "951",
	number = "1",
	pages = "L8",
	year = "2023"
}

@article{Reardon:2023gzh,
	author = "Reardon, Daniel J. and others",
	title = "{Search for an Isotropic Gravitational-wave Background with the Parkes Pulsar Timing Array}",
	eprint = "2306.16215",
	archivePrefix = "arXiv",
	primaryClass = "astro-ph.HE",
	doi = "10.3847/2041-8213/acdd02",
	journal = "Astrophys. J. Lett.",
	volume = "951",
	number = "1",
	pages = "L6",
	year = "2023"
}

@article{EPTA:2023fyk,
	author = "Antoniadis, J. and others",
	collaboration = "EPTA, InPTA:",
	title = "{The second data release from the European Pulsar Timing Array - III. Search for gravitational wave signals}",
	eprint = "2306.16214",
	archivePrefix = "arXiv",
	primaryClass = "astro-ph.HE",
	doi = "10.1051/0004-6361/202346844",
	journal = "Astron. Astrophys.",
	volume = "678",
	pages = "A50",
	year = "2023"
}

@article{Xu:2023wog,
	author = "Xu, Heng and others",
	title = "{Searching for the Nano-Hertz Stochastic Gravitational Wave Background with the Chinese Pulsar Timing Array Data Release I}",
	eprint = "2306.16216",
	archivePrefix = "arXiv",
	primaryClass = "astro-ph.HE",
	doi = "10.1088/1674-4527/acdfa5",
	journal = "Res. Astron. Astrophys.",
	volume = "23",
	number = "7",
	pages = "075024",
	year = "2023"
}

@article{Nester:1998mp,
	author = "Nester, James M. and Yo, Hwei-Jang",
	title = "{Symmetric teleparallel general relativity}",
	eprint = "gr-qc/9809049",
	archivePrefix = "arXiv",
	reportNumber = "NCU-CCS-980904",
	journal = "Chin. J. Phys.",
	volume = "37",
	pages = "113",
	year = "1999"
}

@article{BeltranJimenez:2019esp,
	author = "Beltr{\'a}n Jim{\'e}nez, Jose and Heisenberg, Lavinia and Koivisto, Tomi S.",
	title = "{The Geometrical Trinity of Gravity}",
	eprint = "1903.06830",
	archivePrefix = "arXiv",
	primaryClass = "hep-th",
	doi = "10.3390/universe5070173",
	journal = "Universe",
	volume = "5",
	number = "7",
	pages = "173",
	year = "2019"
}

@article{Heisenberg:2023lru,
	author = "Heisenberg, Lavinia",
	title = "{Review on f(Q) gravity}",
	eprint = "2309.15958",
	archivePrefix = "arXiv",
	primaryClass = "gr-qc",
	doi = "10.1016/j.physrep.2024.02.001",
	journal = "Phys. Rept.",
	volume = "1066",
	pages = "1--78",
	year = "2024"
}

@article{Lymperis:2022oyo,
	author = "Lymperis, Andreas",
	title = "{Late-time cosmology with phantom dark-energy in f(Q) gravity}",
	eprint = "2207.10997",
	archivePrefix = "arXiv",
	primaryClass = "gr-qc",
	doi = "10.1088/1475-7516/2022/11/018",
	journal = "JCAP",
	volume = "11",
	pages = "018",
	year = "2022"
}

@article{Paul:2022jnq,
	author = "Paul, B. C. and Chanda, A. and Beesham, A. and Maharaj, S. D.",
	title = "{Late time cosmology in -gravity with interacting fluids}",
	doi = "10.1088/1361-6382/ac4b97",
	journal = "Class. Quant. Grav.",
	volume = "39",
	number = "6",
	pages = "065006",
	year = "2022"
}

@article{Narawade:2023tnn,
	author = "Narawade, S. A. and Singh, Shashank P. and Mishra, B.",
	title = "{Accelerating cosmological models in f(Q) gravity and the phase space analysis}",
	eprint = "2303.06427",
	archivePrefix = "arXiv",
	primaryClass = "gr-qc",
	doi = "10.1016/j.dark.2023.101282",
	journal = "Phys. Dark Univ.",
	volume = "42",
	pages = "101282",
	year = "2023"
}

@article{Narawade:2023rip,
	author = "Narawade, S. A. and Shekh, S. H. and Mishra, B. and Khyllep, Wompherdeiki and Dutta, Jibitesh",
	title = "{Modelling the accelerating universe with f(Q) gravity: observational consistency}",
	eprint = "2303.01985",
	archivePrefix = "arXiv",
	primaryClass = "gr-qc",
	doi = "10.1140/epjc/s10052-024-13150-5",
	journal = "Eur. Phys. J. C",
	volume = "84",
	number = "8",
	pages = "773",
	year = "2024"
}

@article{Dimakis:2023uib,
	author = "Dimakis, N. and Roumeliotis, M. and Paliathanasis, A. and Christodoulakis, T.",
	title = "{Anisotropic solutions in symmetric teleparallel $f\left( Q\right) $-theory: Kantowski{\textendash}Sachs and Bianchi III LRS cosmologies}",
	eprint = "2304.04419",
	archivePrefix = "arXiv",
	primaryClass = "gr-qc",
	doi = "10.1140/epjc/s10052-023-11964-3",
	journal = "Eur. Phys. J. C",
	volume = "83",
	number = "9",
	pages = "794",
	year = "2023"
}

@article{Sokoliuk:2023ccw,
	author = "Sokoliuk, Oleksii and Arora, Simran and Praharaj, Subhrat and Baransky, Alexander and Sahoo, P. K.",
	title = "{On the impact of f(Q) gravity on the large scale structure}",
	eprint = "2303.17341",
	archivePrefix = "arXiv",
	primaryClass = "astro-ph.CO",
	doi = "10.1093/mnras/stad968",
	journal = "Mon. Not. Roy. Astron. Soc.",
	volume = "522",
	number = "1",
	pages = "252--267",
	year = "2023"
}

@article{Milgrom:2019rtd,
	author = "Milgrom, Mordehai",
	title = "{Noncovariance at low accelerations as a route to MOND}",
	eprint = "1908.01691",
	archivePrefix = "arXiv",
	primaryClass = "gr-qc",
	doi = "10.1103/PhysRevD.100.084039",
	journal = "Phys. Rev. D",
	volume = "100",
	number = "8",
	pages = "084039",
	year = "2019"
}

@article{DAmbrosio:2020nev,
	author = "D'Ambrosio, Fabio and Garg, Mudit and Heisenberg, Lavinia",
	title = "{Non-linear extension of non-metricity scalar for MOND}",
	eprint = "2004.00888",
	archivePrefix = "arXiv",
	primaryClass = "gr-qc",
	doi = "10.1016/j.physletb.2020.135970",
	journal = "Phys. Lett. B",
	volume = "811",
	pages = "135970",
	year = "2020"
}

@article{Bajardi:2020fxh,
	author = "Bajardi, Francesco and Vernieri, Daniele and Capozziello, Salvatore",
	title = "{Bouncing Cosmology in f(Q) Symmetric Teleparallel Gravity}",
	eprint = "2011.01248",
	archivePrefix = "arXiv",
	primaryClass = "gr-qc",
	doi = "10.1140/epjp/s13360-020-00918-3",
	journal = "Eur. Phys. J. Plus",
	volume = "135",
	number = "11",
	pages = "912",
	year = "2020"
}

@article{Agrawal:2021rur,
	author = "Agrawal, A. S. and Pati, Laxmipriya and Tripathy, S. K. and Mishra, B.",
	title = "{Matter bounce scenario and the dynamical aspects in f(Q,T) gravity}",
	eprint = "2108.02575",
	archivePrefix = "arXiv",
	primaryClass = "gr-qc",
	doi = "10.1016/j.dark.2021.100863",
	journal = "Phys. Dark Univ.",
	volume = "33",
	pages = "100863",
	year = "2021"
}

@article{Gadbail:2023loj,
	author = "Gadbail, Gaurav N. and Kolhatkar, Ameya and Mandal, Sanjay and Sahoo, P. K.",
	title = "{Correction to Lagrangian for bouncing cosmologies in f(Q) gravity}",
	eprint = "2304.10245",
	archivePrefix = "arXiv",
	primaryClass = "gr-qc",
	doi = "10.1140/epjc/s10052-023-11798-z",
	journal = "Eur. Phys. J. C",
	volume = "83",
	number = "7",
	pages = "595",
	year = "2023"
}

@article{Dimakis:2021gby,
	author = "Dimakis, N. and Paliathanasis, A. and Christodoulakis, T.",
	title = "{Quantum cosmology in f(Q) theory}",
	eprint = "2108.01970",
	archivePrefix = "arXiv",
	primaryClass = "gr-qc",
	doi = "10.1088/1361-6382/ac2b09",
	journal = "Class. Quant. Grav.",
	volume = "38",
	number = "22",
	pages = "225003",
	year = "2021"
}

@article{Bajardi:2023vcc,
	author = "Bajardi, Francesco and Capozziello, Salvatore",
	title = "{Minisuperspace quantum cosmology in f(Q) gravity}",
	eprint = "2305.00318",
	archivePrefix = "arXiv",
	primaryClass = "gr-qc",
	doi = "10.1140/epjc/s10052-023-11703-8",
	journal = "Eur. Phys. J. C",
	volume = "83",
	number = "6",
	pages = "531",
	year = "2023"
}

@article{Dialektopoulos:2019mtr,
	author = "Dialektopoulos, Konstantinos F. and Koivisto, Tomi S. and Capozziello, Salvatore",
	title = "{Noether symmetries in Symmetric Teleparallel Cosmology}",
	eprint = "1905.09019",
	archivePrefix = "arXiv",
	primaryClass = "gr-qc",
	reportNumber = "NORDITA 2019-048",
	doi = "10.1140/epjc/s10052-019-7106-8",
	journal = "Eur. Phys. J. C",
	volume = "79",
	number = "7",
	pages = "606",
	year = "2019"
}

@article{Ayuso:2020dcu,
	author = "Ayuso, Ismael and Lazkoz, Ruth and Salzano, Vincenzo",
	title = "{Observational constraints on cosmological solutions of $f(Q)$ theories}",
	eprint = "2012.00046",
	archivePrefix = "arXiv",
	primaryClass = "astro-ph.CO",
	doi = "10.1103/PhysRevD.103.063505",
	journal = "Phys. Rev. D",
	volume = "103",
	number = "6",
	pages = "063505",
	year = "2021"
}

@article{Barros:2020bgg,
	author = "Barros, Bruno J. and Barreiro, Tiago and Koivisto, Tomi and Nunes, Nelson J.",
	title = "{Testing $F(Q)$ gravity with redshift space distortions}",
	eprint = "2004.07867",
	archivePrefix = "arXiv",
	primaryClass = "gr-qc",
	doi = "10.1016/j.dark.2020.100616",
	journal = "Phys. Dark Univ.",
	volume = "30",
	pages = "100616",
	year = "2020"
}

@article{Frusciante:2021sio,
	author = "Frusciante, Noemi",
	title = "{Signatures of $f(Q)$-gravity in cosmology}",
	eprint = "2101.09242",
	archivePrefix = "arXiv",
	primaryClass = "astro-ph.CO",
	doi = "10.1103/PhysRevD.103.044021",
	journal = "Phys. Rev. D",
	volume = "103",
	number = "4",
	pages = "044021",
	year = "2021"
}

@article{Aggarwal:2022eae,
	author = "Aggarwal, Nakul and Pourmand, Ali and Shojai, Fatimah and Parthasarathy, Harish",
	title = "{Constraining Generalized Chaplygin Gas in Non-Minimally Coupled $f(Q)$ Cosmology using Quasars and $H(z)$ Data}",
	eprint = "2212.00312",
	archivePrefix = "arXiv",
	primaryClass = "gr-qc",
	month = "12",
	year = "2022"
}

@article{De:2022jvo,
	author = "De, Avik and Loo, Tee-How",
	title = "{On the viability of f(Q) gravity models}",
	eprint = "2212.08304",
	archivePrefix = "arXiv",
	primaryClass = "gr-qc",
	doi = "10.1088/1361-6382/accef7",
	journal = "Class. Quant. Grav.",
	volume = "40",
	number = "11",
	pages = "115007",
	year = "2023"
}

@article{Albuquerque:2022eac,
	author = "Albuquerque, In{\^e}s S. and Frusciante, Noemi",
	title = "{A designer approach to f(Q) gravity and cosmological implications}",
	eprint = "2202.04637",
	archivePrefix = "arXiv",
	primaryClass = "astro-ph.CO",
	doi = "10.1016/j.dark.2022.100980",
	journal = "Phys. Dark Univ.",
	volume = "35",
	pages = "100980",
	year = "2022"
}

@article{Ferreira:2022jcd,
	author = "Ferreira, Jos{\'e} and Barreiro, Tiago and Mimoso, Jos{\'e} and Nunes, Nelson J.",
	title = "{Forecasting F(Q) cosmology with {\ensuremath{\Lambda}}CDM background using standard sirens}",
	eprint = "2203.13788",
	archivePrefix = "arXiv",
	primaryClass = "astro-ph.CO",
	doi = "10.1103/PhysRevD.105.123531",
	journal = "Phys. Rev. D",
	volume = "105",
	number = "12",
	pages = "123531",
	year = "2022"
}

@article{Koussour:2023rly,
	author = "Koussour, M. and De, Avik",
	title = "{Observational constraints on two cosmological models of f(Q) theory}",
	eprint = "2304.11765",
	archivePrefix = "arXiv",
	primaryClass = "gr-qc",
	doi = "10.1140/epjc/s10052-023-11547-2",
	journal = "Eur. Phys. J. C",
	volume = "83",
	number = "5",
	pages = "400",
	year = "2023"
}

@article{Najera:2023wcw,
	author = "N{\'a}jera, Jos{\'e} Antonio and Alvarado, Carlos Ar{\'a}oz and Escamilla-Rivera, Celia",
	title = "{Constraints on f{\,}(Q) logarithmic model using gravitational wave standard sirens}",
	eprint = "2304.12601",
	archivePrefix = "arXiv",
	primaryClass = "gr-qc",
	doi = "10.1093/mnras/stad2180",
	journal = "Mon. Not. Roy. Astron. Soc.",
	volume = "524",
	number = "4",
	pages = "5280--5290",
	year = "2023"
}

@article{Bouali:2023uik,
	author = "Bouali, Amine and Shukla, B. K. and Chaudhary, Himanshu and Tiwari, Rishi Kumar and Samar, Mahvish and Mustafa, G.",
	title = "{Cosmological tests of parametrization q = {\ensuremath{\alpha}} {\ensuremath{-}} {\ensuremath{\beta}} H in f(Q) FLRW cosmology}",
	doi = "10.1142/S0219887823501529",
	journal = "Int. J. Geom. Meth. Mod. Phys.",
	volume = "20",
	number = "09",
	pages = "2350152",
	year = "2023"
}

@article{Ferreira:2023tat,
	author = "Ferreira, Jos{\'e}",
	title = "{Constraining f(Q) Cosmology with Standard Sirens}",
	eprint = "2303.12674",
	archivePrefix = "arXiv",
	primaryClass = "astro-ph.CO",
	month = "3",
	year = "2023"
}

@article{Subramaniam:2023old,
	author = "Subramaniam, Ganesh and De, Avik and Loo, Tee-How and Goh, Yong Kheng",
	title = "{Energy condition bounds on f(Q) model parameters in a curved FLRW Universe}",
	eprint = "2304.05031",
	archivePrefix = "arXiv",
	primaryClass = "gr-qc",
	doi = "10.1016/j.dark.2023.101243",
	journal = "Phys. Dark Univ.",
	volume = "41",
	pages = "101243",
	year = "2023"
}

@article{Capozziello:2023vne,
	author = "Capozziello, Salvatore and De Falco, Vittorio and Ferrara, Carmen",
	title = "{The role of the boundary term in f(Q,~B) symmetric teleparallel gravity}",
	eprint = "2307.13280",
	archivePrefix = "arXiv",
	primaryClass = "gr-qc",
	doi = "10.1140/epjc/s10052-023-12072-y",
	journal = "Eur. Phys. J. C",
	volume = "83",
	number = "10",
	pages = "915",
	year = "2023"
}

@article{De:2023xua,
	author = "De, Avik and Loo, Tee-How and Saridakis, Emmanuel N.",
	title = "{Non-metricity with boundary terms: {\ensuremath{\mathsf{f}}}({\ensuremath{\mathsf{Q}}},{\ensuremath{\mathsf{C}}}) gravity and cosmology}",
	eprint = "2308.00652",
	archivePrefix = "arXiv",
	primaryClass = "gr-qc",
	doi = "10.1088/1475-7516/2024/03/050",
	journal = "JCAP",
	volume = "03",
	pages = "050",
	year = "2024"
}

@article{Paliathanasis:2023pqp,
	author = "Paliathanasis, A. and Dimakis, N. and Christodoulakis, T.",
	title = "{Minisuperspace description of f(Q)-cosmology}",
	eprint = "2308.15207",
	archivePrefix = "arXiv",
	primaryClass = "gr-qc",
	doi = "10.1016/j.dark.2023.101410",
	journal = "Phys. Dark Univ.",
	volume = "43",
	pages = "101410",
	year = "2024"
}

@article{Jarv:2018bgs,
	author = {J{\"a}rv, Laur and R{\"u}nkla, Mihkel and Saal, Margus and Vilson, Ott},
	title = "{Nonmetricity formulation of general relativity and its scalar-tensor extension}",
	eprint = "1802.00492",
	archivePrefix = "arXiv",
	primaryClass = "gr-qc",
	doi = "10.1103/PhysRevD.97.124025",
	journal = "Phys. Rev. D",
	volume = "97",
	number = "12",
	pages = "124025",
	year = "2018"
}

@article{Harko:2018gxr,
	author = "Harko, Tiberiu and Koivisto, Tomi S. and Lobo, Francisco S. N. and Olmo, Gonzalo J. and Rubiera-Garcia, Diego",
	title = "{Coupling matter in modified $Q$ gravity}",
	eprint = "1806.10437",
	archivePrefix = "arXiv",
	primaryClass = "gr-qc",
	doi = "10.1103/PhysRevD.98.084043",
	journal = "Phys. Rev. D",
	volume = "98",
	number = "8",
	pages = "084043",
	year = "2018"
}

@article{Banerjee:2021mqk,
	author = "Banerjee, Ayan and Pradhan, Anirudh and Tangphati, Takol and Rahaman, Farook",
	title = "{Wormhole geometries in $f(Q)$ gravity and the energy conditions}",
	eprint = "2109.15105",
	archivePrefix = "arXiv",
	primaryClass = "gr-qc",
	doi = "10.1140/epjc/s10052-021-09854-7",
	journal = "Eur. Phys. J. C",
	volume = "81",
	number = "11",
	pages = "1031",
	year = "2021"
}

@article{Mustafa:2021bfs,
	author = "Mustafa, G. and Hassan, Zinnat and Sahoo, P. K.",
	title = "{Traversable wormhole inspired by non-commutative geometries in f(Q) gravity with conformal symmetry}",
	eprint = "2112.15112",
	archivePrefix = "arXiv",
	primaryClass = "gr-qc",
	doi = "10.1016/j.aop.2021.168751",
	journal = "Annals Phys.",
	volume = "437",
	pages = "168751",
	year = "2022"
}

@article{Parsaei:2022wnu,
	author = "Parsaei, Foad and Rastgoo, Sara and Sahoo, P. K.",
	title = "{Wormhole in f(Q) gravity}",
	eprint = "2203.06374",
	archivePrefix = "arXiv",
	primaryClass = "gr-qc",
	doi = "10.1140/epjp/s13360-022-03298-y",
	journal = "Eur. Phys. J. Plus",
	volume = "137",
	number = "9",
	pages = "1083",
	year = "2022"
}

@article{Hassan:2022jgn,
	author = "Hassan, Zinnat and Mustafa, G. and Santos, Joao R. L. and Sahoo, P. K.",
	title = "{Embedding procedure and wormhole solutions in f(Q) gravity}",
	eprint = "2207.05304",
	archivePrefix = "arXiv",
	primaryClass = "gr-qc",
	doi = "10.1209/0295-5075/ac8017",
	journal = "EPL",
	volume = "139",
	number = "3",
	pages = "39001",
	year = "2022"
}

@article{Hassan:2022ibc,
	author = "Hassan, Zinnat and Ghosh, Sayantan and Sahoo, P. K. and Rao, V. Sree Hari",
	title = "{GUP corrected Casimir wormholes in f(Q) gravity}",
	eprint = "2209.02704",
	archivePrefix = "arXiv",
	primaryClass = "gr-qc",
	doi = "10.1007/s10714-023-03139-y",
	journal = "Gen. Rel. Grav.",
	volume = "55",
	number = "8",
	pages = "90",
	year = "2023"
}

@article{Hassan:2022hcb,
	author = "Hassan, Zinnat and Ghosh, Sayantan and Sahoo, P. K. and Bamba, Kazuharu",
	title = "{Casimir wormholes in modified symmetric teleparallel gravity}",
	eprint = "2207.09945",
	archivePrefix = "arXiv",
	primaryClass = "gr-qc",
	doi = "10.1140/epjc/s10052-022-11107-0",
	journal = "Eur. Phys. J. C",
	volume = "82",
	number = "12",
	pages = "1116",
	year = "2022"
}

@article{Venkatesha:2023tay,
	author = "Venkatesha, V. and Chalavadi, Chaitra Chooda and Kavya, N. S. and Sahoo, P. K.",
	title = "{Wormhole geometry and three-dimensional embedding in extended symmetric teleparallel gravity}",
	eprint = "2308.07862",
	archivePrefix = "arXiv",
	primaryClass = "gr-qc",
	doi = "10.1016/j.newast.2023.102090",
	journal = "New Astron.",
	volume = "105",
	pages = "102090",
	year = "2024"
}

@article{Jan:2023djj,
	author = {Jan, Munsif and Ashraf, Asifa and Basit, Abdul and Caliskan, Aylin and G{\"u}dekli, Ertan},
	title = "{Traversable Wormhole in $f(Q)$ Gravity Using Conformal Symmetry}",
	doi = "10.3390/sym15040859",
	journal = "Symmetry",
	volume = "15",
	number = "4",
	pages = "859",
	year = "2023"
}

@article{Godani:2023nep,
	author = "Godani, Nisha",
	title = "{Stable traversable wormholes in f(Q) gravity}",
	doi = "10.1142/S0219887823501281",
	journal = "Int. J. Geom. Meth. Mod. Phys.",
	volume = "20",
	number = "08",
	pages = "2350128",
	year = "2023"
}

@article{Javed:2023vmb,
	author = "Javed, Faisal and Mustafa, G. and Mumtaz, Saadia and Atamurotov, Farruh",
	title = "{Thermal analysis with emission energy of perturbed black hole in f(Q) gravity}",
	doi = "10.1016/j.nuclphysb.2023.116180",
	journal = "Nucl. Phys. B",
	volume = "990",
	pages = "116180",
	year = "2023"
}

@article{Mishra:2023bfe,
	author = "Mishra, Ambuj Kumar and Shweta and Sharma, Umesh Kumar",
	title = "{Yukawa{\textendash}Casimir Wormholes in f(Q) Gravity}",
	eprint = "2303.04641",
	archivePrefix = "arXiv",
	primaryClass = "physics.gen-ph",
	doi = "10.3390/universe9040161",
	journal = "Universe",
	volume = "9",
	number = "4",
	pages = "161",
	year = "2023"
}

@article{Chanda:2022cod,
	author = "Chanda, A. and Paul, B. C.",
	title = "{Evolution of primordial black holes in $f(Q)$ gravity with non-linear equation of state}",
	doi = "10.1140/epjc/s10052-022-10579-4",
	journal = "Eur. Phys. J. C",
	volume = "82",
	number = "7",
	pages = "616",
	year = "2022"
}

@article{Wang:2021zaz,
	author = "Wang, Wenyi and Chen, Hua and Katsuragawa, Taishi",
	title = "{Static and spherically symmetric solutions in f(Q) gravity}",
	eprint = "2110.13565",
	archivePrefix = "arXiv",
	primaryClass = "gr-qc",
	doi = "10.1103/PhysRevD.105.024060",
	journal = "Phys. Rev. D",
	volume = "105",
	number = "2",
	pages = "024060",
	year = "2022"
}

@article{Maurya:2022vsn,
	author = "Maurya, S. K. and Mustafa, G. and Govender, M. and Newton Singh, Ksh.",
	title = "{Exploring physical properties of minimally deformed strange star model and constraints on maximum mass limit in f({\ensuremath{\mathscr{Q}}}) gravity}",
	eprint = "2207.02021",
	archivePrefix = "arXiv",
	primaryClass = "gr-qc",
	doi = "10.1088/1475-7516/2022/10/003",
	journal = "JCAP",
	volume = "10",
	pages = "003",
	year = "2022"
}

@article{Maurya:2022wwa,
	author = "Maurya, S. K. and Newton Singh, Ksh. and Lohakare, Santosh V. and Mishra, B.",
	title = "{Anisotropic Strange Star Model Beyond Standard Maximum Mass Limit by Gravitational Decoupling in f(Q)$f(Q)$ Gravity}",
	eprint = "2208.04735",
	archivePrefix = "arXiv",
	primaryClass = "gr-qc",
	doi = "10.1002/prop.202200061",
	journal = "Fortsch. Phys.",
	volume = "70",
	number = "11",
	pages = "2200061",
	year = "2022"
}

@article{Errehymy:2022gws,
	author = "Errehymy, Abdelghani and Ditta, Allah and Mustafa, G. and Maurya, S. K. and Abdel-Aty, Abdel-Haleem",
	title = "{Anisotropic electrically charged stars in f(Q) symmetric teleparallel gravity}",
	doi = "10.1140/epjp/s13360-022-03458-0",
	journal = "Eur. Phys. J. Plus",
	volume = "137",
	number = "12",
	pages = "1311",
	year = "2022"
}

@article{Sokoliuk:2022bwi,
	author = "Sokoliuk, Oleksii and Pradhan, Sneha and Sahoo, P. K. and Baransky, Alexander",
	title = "{Buchdahl quark stars within $f(Q)$ theory}",
	eprint = "2209.11590",
	archivePrefix = "arXiv",
	primaryClass = "gr-qc",
	doi = "10.1140/epjp/s13360-022-03273-7",
	journal = "Eur. Phys. J. Plus",
	volume = "137",
	number = "9",
	pages = "1077",
	year = "2022"
}

@article{Calza:2022mwt,
	author = "Calz{\'a}, Marco and Sebastiani, Lorenzo",
	title = "{A class of static spherically symmetric solutions in f(Q)-gravity}",
	eprint = "2208.13033",
	archivePrefix = "arXiv",
	primaryClass = "gr-qc",
	doi = "10.1140/epjc/s10052-023-11393-2",
	journal = "Eur. Phys. J. C",
	volume = "83",
	number = "3",
	pages = "247",
	year = "2023"
}

@article{Ditta:2023xhx,
	author = "Ditta, Allah and Tiecheng, Xia and Errehymy, Abdelghani and Mustafa, G. and Maurya, S. K.",
	title = "{Anisotropic charged stellar models with modified Van der Waals EoS in f(Q) gravity}",
	doi = "10.1140/epjc/s10052-023-11390-5",
	journal = "Eur. Phys. J. C",
	volume = "83",
	number = "3",
	pages = "254",
	year = "2023"
}

\end{document}